\documentclass[aps,prl,twocolumn,groupedaddress]{revtex4}

\usepackage[dvips]{graphicx}
\usepackage{latexsym,amsfonts,amssymb}
\usepackage{amsmath}

\usepackage{helvet}
\usepackage[dvips,dvipsnames,usenames]{color}
\usepackage{color}

\newcommand{\be}{\begin{equation}}
\newcommand{\ee}{\end{equation}}
\newcommand{\bd}{\begin{displaymath}}
\newcommand{\ed}{\end{displaymath}}

\newcommand{\beq}{\begin{eqnarray}}
\newcommand{\eeq}{\end{eqnarray}}

\newcommand{\HH}{{\cal H}}

\newcommand{\p}{\partial}

\newcommand{\lan}{\langle}
\newcommand{\ran}{\rangle}
\newcommand{\lb}{\left[}
\newcommand{\rb}{\right]}
\newcommand{\lp}{\left(}
\newcommand{\rp}{\right)}
\renewcommand{\phi}{\varphi}

\begin{document}

\title{Adiabatic nonlinear probes of one-dimensional Bose gases}

\author{C. De Grandi}
\email[]{degrandi@bu.edu}
\author{R. A. Barankov}
\author{A. Polkovnikov}

\affiliation{Department of Physics, Boston University, Boston, Massachusetts 02215, USA}

\begin{abstract}

We discuss two complimentary problems: adiabatic loading of one-dimensional bosons into an optical lattice and merging two one-dimensional Bose systems. Both problems can be mapped to the sine-Gordon model. This mapping allows us to find power-law scalings for the number of excitations with the ramping rate in the regime where the conventional linear response approach fails. We show that the exponent of this power law is sensitive to the interaction strength. In particular, the response is larger, or less adiabatic, for strongly (weakly) interacting bosons for the loading (merging) problem. Our results illustrate that in general the nonlinear response to slow relevant perturbations can be a powerful tool for characterizing properties of interacting systems.

\end{abstract}
\maketitle

Dynamics of one-dimensional (1D) cold gases~\cite{ketterlefirst1dbec,greinerbloch1dbec,esslinger1dbec} is characterized by a strong response to external perturbations~\cite{schmiedmayer,newtoncradle} compared to higher-dimensional systems, due to the large density of low-energy states. For example, for slow processes the excitations are strongly enhanced and can qualitatively affect the dynamics and lead to the breakdown of adiabatic approximation~\cite{tolianature}.  Understanding slow dynamics has two-fold implications. First, such dynamics can be used as a probe of interacting systems in the regimes where the linear response (Kubo-type) analysis fails. Second, this understanding can help to devise optimal ways for minimizing heating in the system during slow processes. This minimization can be important for various applications including adiabatic preparation of strongly correlated systems and adiabatic quantum computing. In this Letter we will demonstrate both such implications using two related problems of loading 1D interacting bosons into an optical lattice and merging (or splitting) two 1D condensates.

The physical parameter that defines the properties of 1D bosonic systems is the ratio of the interaction and kinetic energies $\gamma=mc/(\hbar^2\rho)$, where $m$ is the mass of a particle, $c$ is the interaction strength~\cite{Olshanii}, and $\rho$ is the density of particles. In the spatially uniform system, the energy spectrum is gapless in the whole range of the parameter $\gamma$, which at low energies is reflected in the Luttinger liquid description~\cite{Haldane}.
In the extreme limit of strong interactions, $\gamma\gg 1$, Tonks-Girardeau (TG) gas~\cite{girardeaus}, the particles behave as hard-core spheres, with the repulsion playing a role similar to the Pauli exclusion principle.
This regime has been recently realized in cold atoms~\cite{weissTGgas,blochTGgas}.

External perturbations can significantly modify the phase structure of bosonic gases.
A bosonic system undergoes the transition from superfluid to Mott-insulator~\cite{fisher,greiner}, driven by the competition of the kinetic and interaction energy: in deep lattices the insulating phase is realized when the interaction energy becomes larger than the tunneling (kinetic) energy. This mean-field picture of the transition is modified in one dimension. If the repulsion between particles is strong enough, an arbitrarily small periodic potential commensurate with the boson density stabilizes the Mott-insulator phase~\cite{buchler,sfmott1d}. At weaker interactions, the perturbation is irrelevant, and the system is gapless.

%%%%%%%%%%%%%%%%%%%%%%%%%%%%%%%%%%%%%%%%%%%%%%%%%%%%%%%%
\begin{figure}
\includegraphics[width=3.5in]{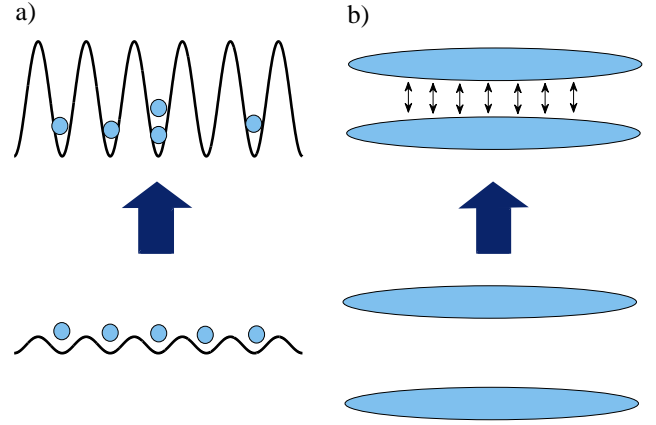}
\caption{
Two processes equivalently described by the sine-Gordon model: a) loading of a 1D Bose gas into a commensurate optical lattice, b) merging of two 1D Bose gases.}
\label{pic1}
\end{figure}
%%%%%%%%%%%%%%%%%%%%%%%%%%%%%%%%%%%%%%%%%%%%%%%%%%%%%%%%%%%%%%%%%

\begin{figure}
\includegraphics[width=3.5in]{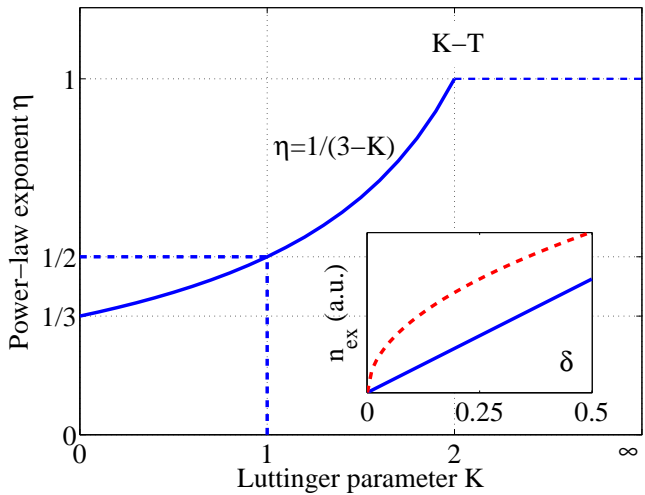}
\caption{Scaling exponent $\eta$ for the excitations $n_{\rm ex} \sim \delta^{\eta}$ as a function of the Luttinger parameter $K$. The Kosterlitz-Thouless (K-T) critical point at $K=2$ separates gapped and gapless phases. The dashed line at $K\ge 2$ indicates linear response to loading of the lattice. The behavior of $\eta$ in the vicinity of $K=2$ is a sketch: a rounding-off of the cusp is expected for a finite system. {\it Inset:}  The total number of excitations as a function of the ramping rate $\delta$ for $K=1$ (dashed red line) and $K\lesssim 2$ (solid blue line).}
\label{exponent}
\vspace{-0.25cm}
\end{figure}

We consider the response of the system to a linear ramp of the lattice amplitude $V(t)$ with time, as it is sketched in Fig.~\ref{pic1}a). A similar problem has been addressed experimentally in the case of a three-dimensional Bose-Einstein condensate~\cite{3Dloading}. The main result of our work is the scaling relation between the total number of excitations and the ramping rate $\delta=|\partial_tV|$:
\beq\label{scaling}
n_{\rm ex} \sim \delta^{\eta}, \quad
\eta=1/(3-K),\quad 0<K<2,
\eeq
where $K$ is the Luttinger parameter. For repulsive bosons
$K\approx 1+4/\gamma$ in the strongly interacting case, $\gamma\gg 1$, and
$K\approx \pi/\sqrt{\gamma}$ for weak interactions, $\gamma\ll 1$~\cite{lieb,cazalilla}. The scaling~(\ref{scaling}) is valid provided that the loading time is much longer than the transition time at which the excitations become effectively frozen (see Eq.(\ref{tr_time}) below).
The scaling exponent $\eta$ increases towards $K=2$, where it saturates to unity. The critical point $K=2$ separates the gapped phase at $K< 2$ where a weak lattice is relevant from the gapless phase at $K>2$ where it is irrelevant.
The response of the system to the perturbation is not universal in the latter case since it depends on high energy cut-off of the effective Luttinger liquid description. Nonetheless, for a finite system, from the linear response approach one expects $\eta=1$ (indicated by the dashed line in Fig.~\ref{exponent}). The relation~(\ref{scaling}) can be generalized to an arbitrary power $r$ of the ramp of the lattice potential in time:
$n_{\rm ex}\propto \delta^{r/(2+r-K)}$.
For the linear ramp $r=1$ this result reduces to Eq.~(\ref{scaling}) and for the exponential ramp $r\to\infty$ we obtain linear scaling as generally expected~\cite{krishnendu, optimal_passage} in 1D systems with the linear spectrum.

%%%%%%%%%%%%%%%%%%%%%%%%%%%%%%%%%%%%%%%%%%%%%%%%%%%
%%%%%%%%%%%%%%%%%%%%%%%%%%%%%%%%%%%%%%%%%%%%%%%%%%%

The scaling relation~(\ref{scaling}) is obtained by mapping our problem to the sine-Gordon (SG) model~\cite{coleman} with time-dependent coupling. The same formalism describes a different system of two identical coupled 1D Bose systems (see ~Fig.~\ref{pic1}b)). The tunneling of particles leads to the Josephson-like energy term depending on the relative phase of the condensates~\cite{toliacoupled}, and this system can be also mapped to the sine-Gordon model.
Merging the condensates by adiabatic increase of the coupling produces defects in the relative phase. This process is characterized by the scaling relation similar to~(\ref{scaling}) with the substitution $K\to 1/(2K)$.
Therefore, contrary to the problem of loading into a lattice, the response of the system is more adiabatic for strongly-interacting bosons. This result is intuitively clear since stronger interactions between bosons lead to larger phase fluctuations and therefore to weaker effect of the tunneling~\cite{giamarchi_ho, ludwig}.
Another system where our results~(\ref{scaling}) apply
is a 1D Bose gas with long-range interactions.
There the transition between the Mott-insulator and the Haldane-insulator phases is also described by the sine-Gordon model~\cite{dallatorre} with the tuning parameter being equivalent to the strength of the periodic potential in the loading problem.

Let us focus on a specific problem of loading 1D bosons into an optical lattice. At low energies the system without the lattice can be described within the Luttinger liquid formalism~\cite{Haldane, cazalilla}. The optical lattice, $V_L(x,t)=V_0(t) \cos(2\pi x/a)$,
%~\cite{note2}
commensurate with the boson density, causes interference between particles moving in opposite directions. These processes generate the well-known sine-Gordon Hamiltonian~\cite{coleman}:
\beq\label{timeSG}
\HH = \frac{1}{2}\int d\,x\, \lb \Phi(x)^2 +(\partial_x \theta)^2-4V(t)\cos(\beta\, \theta)\rb,
\eeq
where the length and the time are measured in units of the lattice spacing $a$ and $a/v_s$, respectively, and the energies are rescaled by a factor $\hbar v_s/a$, with the sound velocity $v_s=v_F/K$ proportional to the Fermi velocity $v_F=\hbar\pi\rho/m$. The conjugate variables $\Phi(x)$ and $\theta(x)$ describe the fluctuations of the superfluid phase and the density, respectively~\cite{cazalilla}. The only important parameter that characterizes the sine-Gordon model is $\beta=2\sqrt{\pi K}$: the gapped phase corresponds to $K<2$ and the gapless does to $K>2$~\cite{coleman}. The spectrum of the gapped phase consists of solitons and anti-solitons
in the so-called repulsive regime at $1\le K<2$, and also their bound states, breathers, in the attractive regime at $0<K<1$~\cite{Zamolodchikovs79}.
Solitons (anti-solitons) have topological charge $+1$ ($-1$), while breathers have zero topological charge.
For repulsive bosons ($K\ge 1$)  it is possible to cross the quantum critical point at $K=2$ upon changing the interaction strength between particles. We note that in a gapless regime, there is a critical depth of the lattice at which a transition to a gapped Mott-insulator state occurs described by the effective renormalized $K=2$~\cite{buchler}. Other models, which we mentioned earlier correspond to other domains of $K$ or equivalently $\beta$.

We assume a linear time-dependence of the lattice amplitude $V(t)=V_f\alpha t$, where $V_f$ is the final value of the lattice, so that the loading time is restricted to $0\le t\le t_f=1/\alpha$. In the gapped regime the excitations are mainly created at initial stage, with their number being effectively unchanged after some characteristic time scale related to the ramping rate $\delta=|V_f\alpha|$ (see below). If the loading time is larger than this scale then our results do not depend on $t_f$ and we are able to make universal predictions.

To study the excitation probability we employ the perturbation theory in the ramping rate~\cite{toliapoint}:
\be\label{eq:exc}
n_{\rm ex}=\sum_{p\ne 0} \left| \int_0^{\infty} dt\, \lan p,t| \p_t| 0,t \ran  e^{i\int_0^{t} dt'\lb \omega_p(t')-\omega_0(t')\rb} \right|^2,
\ee
where we assumed that the system is initially in its ground state, and the upper limit of integration over the loading time was set to infinity. Here $\omega_p(t)-\omega_0(t)$ is the excitation energy defined in the instantaneous basis $|p,t\ran$ with respect to the ground state $|0,t\ran$.
The fact that the matrix element of the perturbation appears in the instantaneous basis highlights the difference of this approach with the conventional linear response, in which unperturbed basis is employed. One can show that in the regimes where our approach gives nontrivial scaling, the linear response in the coupling $V$ diverges.

The time-dependent perturbation in Eq.~(\ref{eq:exc}) conserves the topological charge and the total momentum. Therefore in the adiabatic process the only relevant excitations are the soliton-antisoliton pairs created from the vacuum state. The instantaneous eigen-state $|p,t\ran$ in Eq.(\ref{matrix_element}) corresponds to a pair with momenta $(p,-p)$, and the energy $\omega_p=2\sqrt{p^2+m_s^2}$ with a gap equal to twice the soliton mass $m_s$. Therefore the matrix element $\lan p,t|\p_t|0,t\ran$ for the time-dependent perturbation can be explicitly written as follows:
\be\label{matrix_element}
\lan p,t|\p_t |0,t\ran= 4\p_t V\frac{\lan 0,t|A_\pm(p) A_\mp(-p ) \cos\beta\theta|0,t\ran}{\omega_0(t)-\omega_p(t)},
\ee
where $A^\dagger_\pm(p)$ is the operator that creates a soliton (antisoliton) state with momentum $p$.
The matrix element appearing in Eq.~(\ref{matrix_element}) is called the form-factor of $\cos(\beta\theta)$, and it is known from the literature (see Ref.~\cite{Lukyanov97}).

The scaling behavior of $n_{\rm ex}(\delta)$ is found from the analysis of the phase factor in Eq.(\ref{eq:exc}). The dependence of the soliton mass $m_s\simeq V^{1/(2-K)}$ on the loading amplitude $V$ extracted from the exact result of Ref.~\cite{Zamolodchikov95}, suggests rescaling the variables:
\be\label{eq:scaled_variables}
\tau=\delta^{1/(3-K)}t,\quad q=\delta^{-1/(3-K)}p,
\ee
which leads to the invariant form of the phase $ \int_0^t dt'\lb\omega_p(t')-\omega_0(t')\rb=\varphi\lp q,\tau\rp $.
Similarly one obtains $dt\lan p,t|\p_t |0,t\ran= g(q\,\tau^{1/(K-2)})dV/V$, where the function $g$ can be expressed through the form-factors of Ref.~\cite{Lukyanov97}.

The scaling relations in Eq.(\ref{eq:scaled_variables}) define the characteristic time $t_{tr}$ when the transitions become suppressed. Indeed the value of the integral over time in Eq.(\ref{eq:exc}) is defined by the time-scale at which the phase factor starts to oscillate rapidly. Thus, we obtain an estimate 
\be\label{tr_time}
t_{tr}\simeq a/v_s(\delta a^2/v_s^2\hbar)^{1/(K-3)}. 
\ee
Note that for $K<2$ the loading time $t_f=1/\alpha$ is much longer than the transition time $t_{tr}$ provided that $\hbar\alpha\ll m_s(V_f)$, which justifies the extension of the upper limit of integration over time in Eq.(\ref{eq:exc}) to infinity. As one approaches the K-T point $K=2$, the loading time $t_f$ that obeys the relation becomes exponentially large. At  $K\ge 2$, the system is gapless $m_s=0$, and the scaling in Eq.(\ref{eq:exc}) is non-universal. Putting these results together we obtain the scaling~(\ref{scaling}) of the excitation probability with the ramping rate.

Similarly we can analyze the problem of merging two condensates with tunneling amplitude changing linearly in time, $J(t)=J_f\alpha t$, with $\delta=|J_f\alpha|$ being the ramping rate. The system is described by the Hamiltonian~(\ref{timeSG}) where $\theta$ is the relative phase of the condensates, with the substitutions: $V\to J$ and $K\to 1/(2K)$. This mapping brings the system into
the attractive regime of the SG-model, where the low-energy excitations consist of breathers. The energy gap in the spectrum of breathers is proportional to the soliton mass, which immediately leads to the same rescaling  for momentum and time as in Eqs.~(\ref{eq:scaled_variables}) with the corresponding substitution of $K$. The number of excitations scales with the ramping rate as
\be
\label{scaling1}
n_{\rm ex}\sim \delta^{\eta'}, \quad \eta'=2K/(6K-1),\quad K\ge 1/4.
\ee
The exponent $\eta'$ decreases as one goes from strongly-interacting to weakly-interacting case, indicating that the effects of non-adiabaticity become stronger in this limit.

In a similar fashion one can analyze other observables, for example, energy added to the system in a cyclic process in which the lattice is adiabatically loaded and unloaded (merging and splitting of two 1D condensates). Then the spectrum in the final state is gapless and the heating can be found as $Q=\sum_p\epsilon_p n_p$, where $\epsilon_p$ is the excitation energy and $n_p$ is the occupation of the energy mode. Using similar scaling arguments it can be shown that $Q$ behaves as a square of the number of excitations, $Q\sim \delta^{2/(3-K)}$ for the problem of loading, and $Q\sim \delta^{4K/(6K-1)}$ for the problem of condensates merging.

%%%%%%%%%%%%%%%%%%%%%%%%%%%%%%%%%%%%%%%%%%%%%%%%%%%%%%%%%%
\begin{figure}
\includegraphics[width=3.2in]{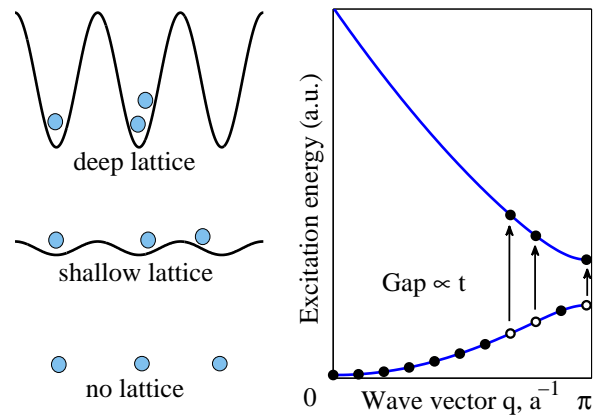}
\caption{Particle-hole excitations in the Tonks-Girardeau limit $K=1$ in real and momentum spaces in the process of loading of the optical lattice. The band-gap opens linearly with time.
\label{bands}}
\end{figure}
%%%%%%%%%%%%%%%%%%%%%%%%%%%%%%%%%%%%%%%%%%%%%%%%%%%%%%%%%%%%%%

We illustrate our main result~(\ref{scaling}) by solving exactly the TG limit $K=1$. In this case, the hard-core bosons are mapped to free spinless fermions~\cite{girardeau}. The effect of the periodic potential is reflected in the momentum space through the presence of a gap between the energy bands. In the case of a unit lattice filling, the gap opens linearly in time between the first and second bands (see Fig.~\ref{bands}). Our calculation of the excitation probability~\cite{inprep} obtained by mapping onto the Landau-Zener problem~\cite{landauzener} leads to the exact number of excitations
\be\label{tg}
n_{\rm ex}=G \frac{m a^2}{\pi^2 \hbar^{3/2}}\,\delta^{1/2},
\ee
where the ramping rate $\delta=|V_f\alpha|$, and $G=0.374$ is an exact numerical factor. Within the perturbative approach~(\ref{eq:exc}), we find the same parametric dependence, except for a slightly bigger numerical factor $G'=0.44$. The constant $G'$ is also reproduced from the form-factor approach using Eqs.~(\ref{eq:exc}) and (\ref{matrix_element}) at $K=1$.

Another exactly solvable limit appears when $\beta\ll 1$ in Eq.~(\ref{timeSG}). Expanding the cosine term we obtain quadratic Hamiltonian with a time-dependent mass term. As a result, we find $n_{\rm ex}\simeq \delta^{1/3} \log{(\delta^{1/3} L)}$ confirming scalings~(\ref{scaling}) and (\ref{scaling1}), in the two limits $K\ll 1$ and  $K\gg 1$, respectively.
An additional factor depending on the system size $L$ indicates breakdown of the adiabatic approximation~\cite{tolianature}.

Our results suggest an interesting observation related to the heating in the process of loading bosons initially prepared in the gapped phase at $K<2$. Instead of loading bosons into the lattice directly in the gapped phase, to minimize the heating, one can choose a different path. For example, since a small lattice commensurate with the boson density is irrelevant perturbation at $K>2$, in the process of loading it can be preferable first to decrease the interaction to values $K\lesssim2$, then load the bosons into the lattice in the gapless phase, and afterwards adiabatically change the coupling to its initial value bringing the system into the gapped phase.

We estimate the number of excitations on loading bosons into an optical lattice by using the exact result~(\ref{tg})
$n_{\rm ex}=0.098 a\sqrt{V_f/(t_f E_r)}$, where $V_f$ is the final height of the optical lattice and $E_r=(\hbar \pi)^2/(2 m a^2)$ is the recoil energy (the lattice spacing $a$ is in $\mu$m and the loading time $t_f=1/\alpha$  in ms). For typical experimental parameters~\cite{3Dloading} $a=0.425{\,\rm \mu m}$ and $V_f/E_r=20$, we obtain $n_{\rm ex}=0.19$ for $t_f=1{\,\rm ms}$, and $n_{\rm ex}=0.06$ for $t_f=10{\,\rm ms}$. While this number can be relatively small in a single ramp process, one can alternatively probe the scaling~(\ref{scaling}) by cyclically loading and unloading bosons in the lattice.

In conclusion, we analyzed zero-temperature adiabatic response of one-dimensional Bose gases in two complimentary settings of loading a gas into an optical lattice and merging of two gases. The number of excitations created in these processes scales as a power-law of the ramping rate with the exponent strongly depending on the interactions in the system (see Eqs.~(\ref{scaling}), (\ref{scaling1})). Our results can be used to probe interactions in the regime where conventional linear response analysis fails.

We acknowledge insightful discussions with  V.~Gritsev and V.~Gurarie, and thank J.~Levinsen for his notes on Landau-Zener problem. We also acknowledge the hospitality of the Institute Henri Poincar\'{e}~ - ~Centre Emilie Borel.
The financial support was provided by CNRS and AFOSR YIP.

\bibliography{mia_fin2}

\begin{thebibliography}{32}
\expandafter\ifx\csname natexlab\endcsname\relax\def\natexlab#1{#1}\fi
\expandafter\ifx\csname bibnamefont\endcsname\relax
  \def\bibnamefont#1{#1}\fi
\expandafter\ifx\csname bibfnamefont\endcsname\relax
  \def\bibfnamefont#1{#1}\fi
\expandafter\ifx\csname citenamefont\endcsname\relax
  \def\citenamefont#1{#1}\fi
\expandafter\ifx\csname url\endcsname\relax
  \def\url#1{\texttt{#1}}\fi
\expandafter\ifx\csname urlprefix\endcsname\relax\def\urlprefix{URL }\fi
\providecommand{\bibinfo}[2]{#2}
\providecommand{\eprint}[2][]{\url{#2}}

\bibitem[{\citenamefont{G\"orlitz~{\it et al.}}(2001)}]{ketterlefirst1dbec}
\bibinfo{author}{\bibfnamefont{A.}~\bibnamefont{G\"orlitz~{\it et al.}}},
  \bibinfo{journal}{Phys. Rev. Lett.} \textbf{\bibinfo{volume}{87}},
  \bibinfo{pages}{130402} (\bibinfo{year}{2001}).

\bibitem[{\citenamefont{Greiner~{\it et al.}}(2001)}]{greinerbloch1dbec}
\bibinfo{author}{\bibfnamefont{M.}~\bibnamefont{Greiner~{\it et al.}}},
  \bibinfo{journal}{Phys. Rev. Lett.} \textbf{\bibinfo{volume}{87}},
  \bibinfo{pages}{160405} (\bibinfo{year}{2001}).

\bibitem[{\citenamefont{Moritz~{\it et al.}}(2003)}]{esslinger1dbec}
\bibinfo{author}{\bibfnamefont{H.}~\bibnamefont{Moritz~{\it et al.}}},
  \bibinfo{journal}{Phys. Rev. Lett.} \textbf{\bibinfo{volume}{91}},
  \bibinfo{pages}{250402} (\bibinfo{year}{2003}).

\bibitem[{\citenamefont{Hofferberth et~al.}(2007)\citenamefont{Hofferberth,
  Lesanovsky, Fischer, Schumm, and Schmiedmayer}}]{schmiedmayer}
\bibinfo{author}{\bibfnamefont{S.}~\bibnamefont{Hofferberth}},
  \bibinfo{author}{\bibfnamefont{I.}~\bibnamefont{Lesanovsky}},
  \bibinfo{author}{\bibfnamefont{B.}~\bibnamefont{Fischer}},
  \bibinfo{author}{\bibfnamefont{T.}~\bibnamefont{Schumm}}, \bibnamefont{and}
  \bibinfo{author}{\bibfnamefont{J.}~\bibnamefont{Schmiedmayer}},
  \bibinfo{journal}{\nat} \textbf{\bibinfo{volume}{449}}, \bibinfo{pages}{324}
  (\bibinfo{year}{2007}).

\bibitem[{\citenamefont{Kinoshita et~al.}(2006)\citenamefont{Kinoshita, Wenger,
  and Weiss}}]{newtoncradle}
\bibinfo{author}{\bibfnamefont{T.}~\bibnamefont{Kinoshita}},
  \bibinfo{author}{\bibfnamefont{T.}~\bibnamefont{Wenger}}, \bibnamefont{and}
  \bibinfo{author}{\bibfnamefont{D.~S.} \bibnamefont{Weiss}},
  \bibinfo{journal}{\nat} \textbf{\bibinfo{volume}{440}}, \bibinfo{pages}{900}
  (\bibinfo{year}{2006}).

\bibitem[{\citenamefont{Polkovnikov and Gritsev}(2008)}]{tolianature}
\bibinfo{author}{\bibfnamefont{A.}~\bibnamefont{Polkovnikov}} \bibnamefont{and}
  \bibinfo{author}{\bibfnamefont{V.}~\bibnamefont{Gritsev}},
  \bibinfo{journal}{Nature Phys.} \textbf{\bibinfo{volume}{4}},
  \bibinfo{pages}{477} (\bibinfo{year}{2008}).

\bibitem[{\citenamefont{Olshanii}(1998)}]{Olshanii}
\bibinfo{author}{\bibfnamefont{M.}~\bibnamefont{Olshanii}},
  \bibinfo{journal}{Phys. Rev. Lett.} \textbf{\bibinfo{volume}{81}},
  \bibinfo{pages}{938} (\bibinfo{year}{1998}).

\bibitem[{\citenamefont{Haldane}(1981)}]{Haldane}
\bibinfo{author}{\bibfnamefont{F.~D.~M.} \bibnamefont{Haldane}},
  \bibinfo{journal}{Phys. Rev. Lett.} \textbf{\bibinfo{volume}{47}},
  \bibinfo{pages}{1840} (\bibinfo{year}{1981}).

\bibitem[{gir()}]{girardeaus}
\bibinfo{note}{M.~Girardeau, J. Math .Phys. {\bf 1}, 516 (1960); C.~N.~Yang and
  Y.~P.~Yang, J. Math .Phys. {\bf 10}, 1115 (1969); L.~Tonks, Phys. Rev. {\bf
  50}, 955 (1936).}

\bibitem[{\citenamefont{Kinoshita et~al.}(2004)\citenamefont{Kinoshita, Wenger,
  and Weiss}}]{weissTGgas}
\bibinfo{author}{\bibfnamefont{T.}~\bibnamefont{Kinoshita}},
  \bibinfo{author}{\bibfnamefont{T.}~\bibnamefont{Wenger}}, \bibnamefont{and}
  \bibinfo{author}{\bibfnamefont{D.~S.} \bibnamefont{Weiss}},
  \bibinfo{journal}{Science} \textbf{\bibinfo{volume}{305}},
  \bibinfo{pages}{1125} (\bibinfo{year}{2004}).

\bibitem[{\citenamefont{Paredes~{\it et al.}}(2004)}]{blochTGgas}
\bibinfo{author}{\bibfnamefont{B.}~\bibnamefont{Paredes~{\it et al.}}},
  \bibinfo{journal}{\nat} \textbf{\bibinfo{volume}{429}}, \bibinfo{pages}{277}
  (\bibinfo{year}{2004}).

\bibitem[{\citenamefont{Fisher~{\it et al.}}(1989)}]{fisher}
\bibinfo{author}{\bibfnamefont{M.~P.~A.} \bibnamefont{Fisher~{\it et al.}}},
  \bibinfo{journal}{Phys. Rev. B} \textbf{\bibinfo{volume}{40}},
  \bibinfo{pages}{546} (\bibinfo{year}{1989}).

\bibitem[{\citenamefont{Greiner~{\it et al.}}(2002)}]{greiner}
\bibinfo{author}{\bibfnamefont{M.}~\bibnamefont{Greiner~{\it et al.}}},
  \bibinfo{journal}{\nat} \textbf{\bibinfo{volume}{415}}, \bibinfo{pages}{39}
  (\bibinfo{year}{2002}).

\bibitem[{\citenamefont{B\"uchler et~al.}(2003)\citenamefont{B\"uchler,
  Blatter, and Zwerger}}]{buchler}
\bibinfo{author}{\bibfnamefont{H.~P.} \bibnamefont{B\"uchler}},
  \bibinfo{author}{\bibfnamefont{G.}~\bibnamefont{Blatter}}, \bibnamefont{and}
  \bibinfo{author}{\bibfnamefont{W.}~\bibnamefont{Zwerger}},
  \bibinfo{journal}{Phys. Rev. Lett.} \textbf{\bibinfo{volume}{90}},
  \bibinfo{pages}{130401} (\bibinfo{year}{2003}).

\bibitem[{\citenamefont{St\"oferle~{\it et al.}}(2004)}]{sfmott1d}
\bibinfo{author}{\bibfnamefont{T.}~\bibnamefont{St\"oferle~{\it et al.}}},
  \bibinfo{journal}{Phys. Rev. Lett.} \textbf{\bibinfo{volume}{92}},
  \bibinfo{pages}{130403} (\bibinfo{year}{2004}).

\bibitem[{\citenamefont{Gericke~{\it et al.}}(2007)}]{3Dloading}
\bibinfo{author}{\bibfnamefont{T.}~\bibnamefont{Gericke~{\it et al.}}},
  \bibinfo{journal}{Journal of Modern Optics} \textbf{\bibinfo{volume}{54}},
  \bibinfo{pages}{735} (\bibinfo{year}{2007}).

\bibitem[{\citenamefont{Lieb and Liniger}(1963)}]{lieb}
\bibinfo{author}{\bibfnamefont{E.~H.} \bibnamefont{Lieb}} \bibnamefont{and}
  \bibinfo{author}{\bibfnamefont{W.}~\bibnamefont{Liniger}},
  \bibinfo{journal}{Phys. Rev.} \textbf{\bibinfo{volume}{130}},
  \bibinfo{pages}{1605} (\bibinfo{year}{1963}).

\bibitem[{\citenamefont{Cazalilla}(2004)}]{cazalilla}
\bibinfo{author}{\bibfnamefont{M.~A.} \bibnamefont{Cazalilla}},
  \bibinfo{journal}{J. Phys. B} \textbf{\bibinfo{volume}{37}},
  \bibinfo{pages}{S1} (\bibinfo{year}{2004}).

\bibitem[{\citenamefont{Sen et~al.}(2008)\citenamefont{Sen, Sengupta, and
  Mondal}}]{krishnendu}
\bibinfo{author}{\bibfnamefont{D.}~\bibnamefont{Sen}},
  \bibinfo{author}{\bibfnamefont{K.}~\bibnamefont{Sengupta}}, \bibnamefont{and}
  \bibinfo{author}{\bibfnamefont{S.}~\bibnamefont{Mondal}},
  \bibinfo{journal}{Phys. Rev. Lett.} \textbf{\bibinfo{volume}{101}},
  \bibinfo{pages}{016806} (\bibinfo{year}{2008}).

\bibitem[{\citenamefont{Barankov and Polkovnikov}(2008)}]{optimal_passage}
\bibinfo{author}{\bibfnamefont{R.}~\bibnamefont{Barankov}} \bibnamefont{and}
  \bibinfo{author}{\bibfnamefont{A.}~\bibnamefont{Polkovnikov}},
  \bibinfo{journal}{Phys. Rev. Lett.} \textbf{\bibinfo{volume}{101}},
  \bibinfo{pages}{076801} (\bibinfo{year}{2008}).

\bibitem[{\citenamefont{Coleman}(1975)}]{coleman}
\bibinfo{author}{\bibfnamefont{S.}~\bibnamefont{Coleman}},
  \bibinfo{journal}{Phys. Rev. D} \textbf{\bibinfo{volume}{11}},
  \bibinfo{pages}{2088} (\bibinfo{year}{1975}).

\bibitem[{\citenamefont{Gritsev et~al.}(2007)\citenamefont{Gritsev,
  Polkovnikov, and Demler}}]{toliacoupled}
\bibinfo{author}{\bibfnamefont{V.}~\bibnamefont{Gritsev}},
  \bibinfo{author}{\bibfnamefont{A.}~\bibnamefont{Polkovnikov}},
  \bibnamefont{and} \bibinfo{author}{\bibfnamefont{E.}~\bibnamefont{Demler}},
  \bibinfo{journal}{Phys. Rev. B} \textbf{\bibinfo{volume}{75}},
  \bibinfo{eid}{174511} (\bibinfo{year}{2007}).

\bibitem[{\citenamefont{Ho et~al.}(2004)\citenamefont{Ho, Cazalilla, and
  Giamarchi}}]{giamarchi_ho}
\bibinfo{author}{\bibfnamefont{A.~F.} \bibnamefont{Ho}},
  \bibinfo{author}{\bibfnamefont{M.~A.} \bibnamefont{Cazalilla}},
  \bibnamefont{and}
  \bibinfo{author}{\bibfnamefont{T.}~\bibnamefont{Giamarchi}},
  \bibinfo{journal}{Phys. Rev. Lett.} \textbf{\bibinfo{volume}{92}},
  \bibinfo{eid}{130405} (\bibinfo{year}{2004}).

\bibitem[{\citenamefont{Mathey et~al.}(2008)\citenamefont{Mathey, Polkovnikov,
  and Neto}}]{ludwig}
\bibinfo{author}{\bibfnamefont{L.}~\bibnamefont{Mathey}},
  \bibinfo{author}{\bibfnamefont{A.}~\bibnamefont{Polkovnikov}},
  \bibnamefont{and} \bibinfo{author}{\bibfnamefont{A.~C.} \bibnamefont{Neto}},
  \bibinfo{journal}{Europhys. Lett.} \textbf{\bibinfo{volume}{81}},
  \bibinfo{eid}{10008} (\bibinfo{year}{2008}).

\bibitem[{dal()}]{dallatorre}
\bibinfo{note}{E. G. Dalla Torre, E. Berg, and E. Altman, Phys. Rev. Lett
  \textbf{97}, 260401 (2006); E. Berg \textit{et al.}, Phys. Rev. B
  \textbf{77}, 245119 (2008)}.

\bibitem[{\citenamefont{Zamolodchikov and
  Zamolodchikov}(1979)}]{Zamolodchikovs79}
\bibinfo{author}{\bibfnamefont{A.~B.} \bibnamefont{Zamolodchikov}}
  \bibnamefont{and} \bibinfo{author}{\bibfnamefont{A.~B.}
  \bibnamefont{Zamolodchikov}}, \bibinfo{journal}{Annals of Physics}
  \textbf{\bibinfo{volume}{120}}, \bibinfo{pages}{253} (\bibinfo{year}{1979}).

\bibitem[{\citenamefont{Polkovnikov}(2005)}]{toliapoint}
\bibinfo{author}{\bibfnamefont{A.}~\bibnamefont{Polkovnikov}},
  \bibinfo{journal}{Phys. Rev. B} \textbf{\bibinfo{volume}{72}},
  \bibinfo{pages}{R161201} (\bibinfo{year}{2005}).

\bibitem[{\citenamefont{Lukyanov and Zamolodchikov}(1997)}]{Lukyanov97}
\bibinfo{author}{\bibfnamefont{S.}~\bibnamefont{Lukyanov}} \bibnamefont{and}
  \bibinfo{author}{\bibfnamefont{A.}~\bibnamefont{Zamolodchikov}},
  \bibinfo{journal}{Nuclear Physics B} \textbf{\bibinfo{volume}{493}},
  \bibinfo{pages}{571} (\bibinfo{year}{1997}).

\bibitem[{\citenamefont{Zamolodchikov}(1995)}]{Zamolodchikov95}
\bibinfo{author}{\bibfnamefont{A.}~\bibnamefont{Zamolodchikov}},
  \bibinfo{journal}{Int. J. Mod. Phys. A} \textbf{\bibinfo{volume}{10}},
  \bibinfo{pages}{1125} (\bibinfo{year}{1995}).

\bibitem[{\citenamefont{Girardeau}(1960)}]{girardeau}
\bibinfo{author}{\bibfnamefont{M.}~\bibnamefont{Girardeau}},
  \bibinfo{journal}{J. Math. Phys.} \textbf{\bibinfo{volume}{1}},
  \bibinfo{pages}{516} (\bibinfo{year}{1960}).

\bibitem[{inp()}]{inprep}
\bibinfo{note}{C. De Grandi, R. A. Barankov and A. Polkovnikov, in
  preparation}.

\bibitem[{lan()}]{landauzener}
\bibinfo{note}{L. Landau, Phys. Zs. Sowjet. {\bf 2}, 46 (1932); C.~Zener, Proc.
  R. Soc. London, Ser. A \textbf{137}, 696 (1932)}.

\end{thebibliography}

\end{document}